\documentclass[authoryear,preprint,review,12pt]{elsarticle}

\usepackage{amssymb}
\usepackage{amsmath}

\usepackage{url}
\usepackage{caption}
\usepackage{ifthen}
\usepackage{xcolor}
\usepackage{hyperref}
\usepackage{booktabs}
\usepackage{tabularx}
\usepackage{color}
\usepackage{xspace}
\usepackage[most]{tcolorbox}

\newboolean{showcomments}
\setboolean{showcomments}{true}         
\ifthenelse{\boolean{showcomments}}
  {\newcommand{\nb}[2]{
  \fbox{\bfseries\sffamily\scriptsize#1}
     {\sf\small$\blacktriangleright$\textit{\textcolor{red}{#2}}$\blacktriangleleft$}
   }
  }
  {\newcommand{\nb}[2]{}
   
  }

\newcommand\new[1]{{\color{black}#1}} 
\newcommand\newt[1]{{\color{black}#1}}

\journal{Journal of Systems and Software}

\begin{document}

\begin{frontmatter}



\title{\new{Cross-Site Scripting Adversarial Attacks Based on Deep Reinforcement Learning: Evaluation and Extension Study}}


\author[usi]{Samuele Pasini}
\author[usi]{Gianluca Maragliano}
\author[usi]{Jinhan Kim}
\author[usi]{Paolo Tonella}

\affiliation[usi]{organization={Universita della Svizzera italiana},
            addressline={Via G. Buffi 13},
            city={Lugano},
            postcode={6900},
            state={Ticino},
            country={Switzerland}}


\begin{abstract}
Cross-site scripting (XSS) poses a significant threat to web application security. While Deep Learning (DL) has shown remarkable success in detecting XSS attacks, it remains vulnerable to adversarial attacks due to the discontinuous nature of \newt{the mapping between the input (i.e., the attack) and the output (i.e., the prediction of the model whether an input is classified as XSS or benign)}. These adversarial attacks employ mutation-based strategies for different components of XSS attack vectors, allowing adversarial agents to iteratively select mutations to evade detection.
Our work replicates a state-of-the-art XSS adversarial attack, highlighting threats to validity in the reference work and extending it towards a more effective evaluation strategy. Moreover, we introduce an XSS Oracle to mitigate these threats. The experimental results show that our approach achieves an escape rate above 96\% when the threats to validity of the replicated technique are addressed. 
\end{abstract}




\end{frontmatter}



\section{Introduction}
\label{sec:introduction}
The proliferation of web applications has brought significant advancements but also introduced new security challenges. Among the various web-based attacks~\cite{10008276, aaa9491117}, Cross-site scripting (XSS)~\cite{RODRIGUEZ2020106960} stands out as one of the most critical concerns. XSS attacks pose a significant threat as they can compromise user data, steal information, and spread worms. Malicious actors exploit vulnerabilities in web applications to inject harmful scripts, which are then unknowingly executed by users' browsers. To mitigate XSS attacks, robust detection methods and strong input validation techniques are essential to safeguard user data and system integrity. 

Researchers have focused on the XSS vulnerability discovery, employing either static or dynamic analysis. 
Static analysis methods scrutinize the source code to identify potential attacks~\cite{doupe2013dedacota,steinhauser2016static,mohammadi2017detecting,kronjee2018discovering}, but their application might not scale to the size of modern web applications or might result in overconservative results, with several false positives, due to the presence of programming constructs that are difficult to handle statically.  
Dynamic analysis, on the other hand, simulates user operations to detect attacks~\cite{lekies2013detection,stock2014precise,fazzini2015automatically}. However, this approach suffers from a high false negative rate as test cases cannot cover all possible scenarios. To address these limitations, researchers have proposed methods to detect the injection of XSS scripts at runtime, complementing XSS vulnerability discovery before release. In the early approaches, machine learning techniques with manual feature extraction were extensively used~\cite{likarish2009obfuscated,nunan2012automatic,wang2014machine,rathore2017xssclassifier,mereani2018howe}, followed by the advent of Deep Learning (DL) and the use of Deep Neural Networks (DNNs) for XSS detection~\cite{fang2018deepxss,mokbal2019mlpxss,tekerek2021novel}.

While DNNs have shown great promise, they are vulnerable to adversarial attacks~\cite{goodfellow2014generative}, where slight changes to input data can deceive the model. These attacks have successfully compromised DNNs used for XSS attack detection as well. A reference paper by Chen et al.~\cite{CHEN2022102831} proposed a Reinforcement Learning (RL) strategy to generate XSS adversarial examples and attack state-of-the-art (SOTA) XSS attack detectors based on DNNs. Their approach involves preprocessing, tokenization, and word vector representation using the Word2Vec model~\cite{mikolov2013efficient}. The authors achieved almost perfect detection results (over 99\% accuracy) and an impressive Escape Rate (ER)\footnote{he ER represents the percentage of adversarial examples that are not detected as malicious by the detector.} of more than 90\% against all DNN-based detectors.

However, we identified several threats to validity in the work by Chen et al.~\cite{CHEN2022102831}. The first threat to validity is that the application of a sequence of actions could deteriorate the characteristics of the XSS script, and the authors did not apply any strategy to evaluate if the applied sequence of mutations is semantically preserving. The second threat is that the preprocessing pipeline of the detectors does not consider any potentially adversarial example, such that a mutation may potentially result in an out-of-vocabulary token (OOV) that is replaced by `None' in the word vector representation. As a consequence, the input  to the detector is  no longer recognized as an XSS. The last threat concerns the lack of availability of different parts of the reference work, leading to a difficult replication, evaluation, and comparison.

In this paper, we replicate Chen et al.~\cite{CHEN2022102831} using a publicly available dataset and introduce an Oracle for XSS to test the occurrence of the hypothesized threats to validity. We extend the approach towards a more effective strategy by integrating the Oracle into the training process, addressing the identified threats while preserving the effectiveness of the original method. The proposed adversarial agent achieved a performance comparable to the reference work (less than 2\% worse) while completely removing the threats to validity (more than 90\% mitigation), demonstrating a more transparent evaluation strategy and a more effective training strategy. 

The technical contributions of this paper are as follows: 
\begin{itemize}
   \item We replicate a reference work on Deep Reinforcement Learning for XSS adversarial attacks, using publicly available data and the public release of results.
   
   \item We identify the threats to the validity of the reference work and propose a method to mitigate them.
   
   \item We extend the reference work towards a more effective evaluation strategy by introducing an XSS Oracle and integrating it into the training process, effectively addressing the identified threats to validity.
\end{itemize}

The rest of the paper is organized as follows. Section~\ref{sec:backgroud} explores the background related to XSS, Reinforcement Learning (RL), and XSS adversarial approaches, which are needed as preliminaries to understand the reference work.
Section~\ref{sec:reference} analyzes the reference work, with a focus on the possible threats to validity.
Section~\ref{sec:method} presents the proposed method, focusing on the usage of an XSS Oracle and its integration into the reference work.
Section~\ref{sec:empirical_study} describes the research questions, the experimental setting, and the process followed to replicate and extend the reference work.
Section~\ref{sec:results} analyzes the results, Section~\ref{sec:threats} describes the threats-to-validity of our work, while Section~\ref{sec:conclusion} concludes the paper.

\section{Background and Related Work}
\label{sec:backgroud}

\subsection{Cross-Site Scripting (XSS)}

Cross-Site Scripting (XSS) consists of the injection of malicious code into a web page. When a user visits the page, their browser unknowingly executes this script, leading to critical security breaches. It has been recognized as one of the prevalent threats, evidenced by the Open Web Application Security Project (OWASP),\footnote{\url{https://owasp.org/www-project-top-ten/}} a renowned authority on web application security, that has consistently ranked XSS as one of the top ten web application security risks. These attacks can have various malicious intentions, such as stealing confidential information or impersonating users to perform unauthorized actions. There are three primary types of XSS attacks, each with its own characteristics:
\begin{enumerate}

    \item \textit{Stored XSS} is the most severe form, where the malicious script is stored permanently in the server-side database. Any user accessing the affected page risks executing this script, potentially affecting multiple users.

    \item \textit{Reflected XSS} is a more targeted attack. The attacker lures the victim into visiting a malicious URL, often through spam emails. The URL contains harmful code, which is then executed in the victim's browser. This type of XSS is temporary and affects a specific user.

    \item \textit{DOM-based XSS} manipulates the Document Object Model (DOM) of a web page by modifying the input. This triggers the attack when the DOM is parsed on the client side, making it another non-persistent form of XSS.

\end{enumerate}

\subsubsection{Defending Against XSS Attacks}

Given the diverse and harmful nature of XSS attacks, researchers have devoted efforts to developing effective defence strategies. The primary research focus has been on two key areas: XSS vulnerability discovery and XSS attack detection. 

\textbf{XSS vulnerability discovery}: This encompasses both static and dynamic analysis techniques. \textit{Static analysis} searches along all the possible execution paths in the code to find potential attacks. Several approaches have been proposed in this category. Doupe et al.~\cite{doupe2013dedacota} suggested a server-side XSS mitigation strategy that isolates code from data, but this method falls short of dynamic JavaScript attacks. Steinhauser et al.~\cite{steinhauser2016static} developed JSPChecker, a tool employing data flow analysis and string parsing to detect vulnerabilities in sanitization sequences. Mohammadi et al.~\cite{mohammadi2017detecting} utilized automated unit testing to identify vulnerabilities arising from improper input data handling. Kronjee et al.~\cite{kronjee2018discovering} applied machine learning with a 79\% precision rate to detect XSS and SQL injection vulnerabilities through static code analysis. \textit{Dynamic analysis}, on the other hand, involves monitoring the data flow to pinpoint injection points and then testing for actual vulnerabilities. Lekies et al.~\cite{lekies2013detection} introduced a technique to detect DOM-based XSS by monitoring and exploiting vulnerabilities in sensitive calls. Fazzini et al.~\cite{fazzini2015automatically} automatically implemented Content Security Policies (CSP) in web applications, to track and manage the dynamic content.

\textbf{XSS attack detection:} While vulnerability discovery is essential, it may not offer complete protection against XSS attacks. Hence, researchers have also developed methods to identify malicious user input at runtime. This task is challenging due to the obfuscation techniques employed by attackers. Consequently, many detection methods rely on ML and DL approaches. Likarish et al.~\cite{likarish2009obfuscated} used JavaScript features for detection, achieving 92\% accuracy. Nunan et al.~\cite{nunan2012automatic} refined this approach, improving detection. Mereani et al.~\cite{mereani2018howe} extracted structural and behavioral features, reaching 99\% accuracy. Fang et al.~\cite{fang2018deepxss} utilized Word2Vec and LSTM, achieving precision and recall of 99.5\% and 98.7\%, respectively. Mokbal et al.~\cite{mokbal2019mlpxss} constructed a large dataset and developed a feature selection technique, attaining 99.32\% accuracy and 98.35\% recall. Tekerek et al.~\cite{tekerek2021novel} employed a CNN, achieving 97.07\% accuracy on a public dataset.
Despite the impressive results, State-Of-The-Art (SOTA) approaches present vulnerabilities that can be exploited by attackers, among which  vulnerabilities to adversarial attacks against ML/DL.

\subsubsection{Adversarial Attacks on XSS Detectors}

The recent emergence of DL has led to groundbreaking advancements in various fields, including XSS attack detection, where it has achieved SOTA performance. However, researchers have identified a critical issue: the susceptibility of these methods to adversarial attacks. These attacks have successfully evaded multiple DL models across different domains, underscoring the imperative to enhance the robustness of these models. \new{In the adversarial ML domain, several techniques are considered, and many of them are gradient-based, since they try to build adversarial examples applying noise computed by backpropagation \cite{goodfellow2014explaining}. Differently, in} the context of XSS attack detectors, several studies have explored adversarial attacks \new{and most of them are based on Reinforcement Learning.} Fang et al.~\cite{fang2018deepxss} developed an XSS adversarial approach utilizing the Dueling Deep Q Networks algorithm, but its escape detection success rate remained below 10\% due to a simplistic bypass strategy. Zhang et al.~\cite{zhang2020adversarial} proposed an algorithm based on Monte Carlo Tree Search (MCTS) to generate XSS adversarial examples for training the detection model. However, this algorithm relied on limited escape strategies and exhibited high time complexity. Wang et al.~\cite{wang2022black} introduced a method employing soft Q-learning, dividing the bypass process into HTML and JS stages, achieving an impressive 85\% escape rate.

The reference work by Chen et al.~\cite{CHEN2022102831} stands out with its Deep Reinforcement Learning algorithm, leveraging a set of mutation rules as actions, resulting in near-perfect escape rates against various SOTA XSS attack detectors. This approach will be thoroughly analyzed in Section~\ref{sec:reference}.

\newt{
Beyond adversarial studies in security domains such as XSS, recent work has examined adversarial attacks against neural code models, particularly pre-trained programming language models. Although these models achieve strong performance on tasks such as code classification and vulnerability prediction, multiple studies have shown them to be vulnerable to semantics-preserving adversarial transformations. \cite{10.1145/3510003.3510146} showed that natural, semantics-preserving code transformations can significantly degrade model performance without altering program behavior. Subsequent work confirmed the vulnerability of neural code models to superficial structural changes~\cite{10.1145/3428230,10.5555/3524938.3525022}. More recent approaches focus on discrete, token-level attacks that operate directly in the program space and enable effective black-box adversarial generation while preserving functional equivalence~\cite{10.1145/3591227,10.1609/aaai.v37i12.26739}. These findings highlight limitations of gradient-based adversarial methods in structured domains, drawing parallels with adversarial attacks on security payloads such as XSS.
}

\subsection{Reinforcement Learning}

Reinforcement Learning (RL) is a distinctive machine learning paradigm that aims to maximize long-term rewards by striking a balance between exploration and exploitation. Unlike supervised learning, RL does not rely on labeled input-output pairs. Instead, it models the learning process as the interaction between two key components: the agent and the environment. The environment is represented as a timed sequence of states, $S=\langle s_0, s_1, \ldots\rangle$. At any given time $t$, the agent observes a state $s_t$ and selects an action $a_t$ from the available action space $A=\{a_0, a_1, \ldots\}$ according to a policy $\pi(a_t | s_t)$, which is either the same being learned during the agent's interactions with the environment (\textit{on-policy} learning) or which is kept separate from the policy under training (\textit{off-policy} learning). The chosen action triggers a state change, and the new state $s_{t+1}$ is determined by a Markov decision process with probability transition matrix $P(s_{t+1} | s_{t}, a_{t})$. Simultaneously, the agent receives a reward $r_{t+1}$. The agent's objective is to maximize the long-term reward $R=\sum_{t=0}^{\infty}{\gamma^t r_t}$, where $\gamma \in [0,1]$ is a discount factor. This balance between immediate and future rewards is a hallmark of RL.

Several algorithms have been developed for RL, each with its unique characteristics. One widely adopted algorithm is Proximal Policy Optimization (PPO)~\cite{schulman2017proximal}, an on-policy algorithm that alternates between data collection through environment interactions and optimization of a clipped surrogate objective function via stochastic gradient descent. The clipping mechanism stabilizes training by limiting the policy updates, preventing drastic changes. Deep Deterministic Policy Gradient (DDPG)~\cite{lillicrap2020continuous} is an off-policy algorithm where the agent learns a deterministic policy guided by a Q-value function critic, which estimates the value of the optimal policy. DDPG employs target actor and critic networks and an experience replay buffer to enhance stability and learning efficiency. Soft Actor-Critic (SAC)~\cite{haarnoja2018soft}, another off-policy algorithm, is based on the maximum entropy framework. SAC trains the actor to maximize both expected reward and entropy, encouraging broader exploration. This approach has been shown to improve learning speed compared to state-of-the-art methods optimizing the traditional RL objective function. DDPG and SAC are typically applied to continuous action spaces. 
In contrast, PPO is versatile, supporting both continuous and discrete action spaces.
RL approaches are very useful in several domains~\cite{SHAKYA2023120495}, including  adversarial attack generation.
The reference work by Chen et al.~\cite{CHEN2022102831} proposed to train an adversarial agent able to attack XSS detectors using RL.
\section{Reference Work}
\label{sec:reference}

In this section, we  introduce the reference work~\cite{CHEN2022102831} as follows. We begin with an overview of the proposed method. Then, we delve into the \new{experiments described in the reference work}. Lastly, we describe the potential threats to validity we identified, which prompted this replication and extension study.

\subsection{Proposed Method} \label{sec:proposed_method}

\begin{figure}[!t]
  \centering
  \includegraphics[width=0.9\columnwidth]{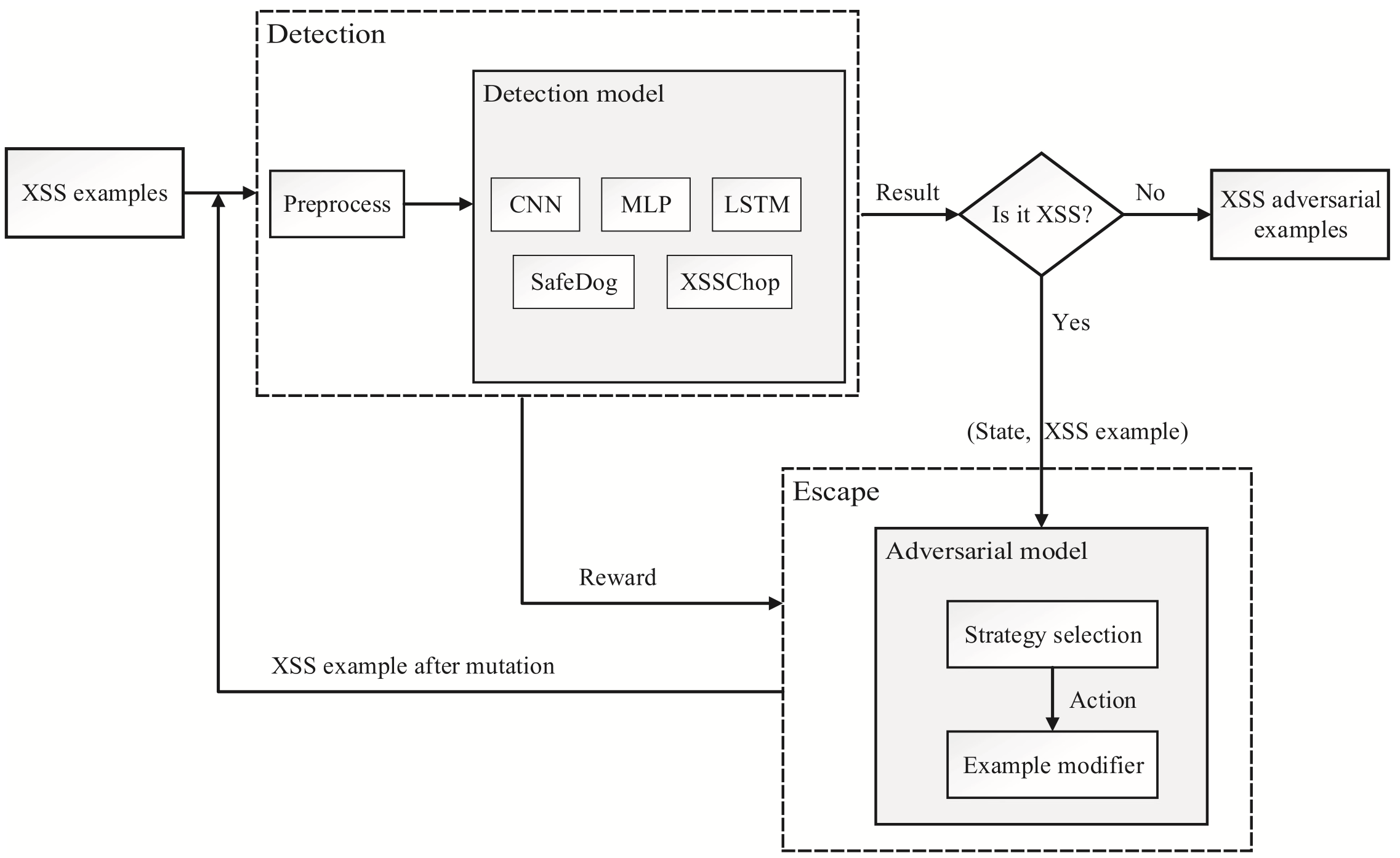}
  
  \caption{Method proposed by the reference work (picture taken from the paper by Chen et al.~\cite{CHEN2022102831})} \label{fig:reference_method}
\end{figure}

As depicted in Figure~\ref{fig:reference_method}, the authors of the reference work introduce a two-stage method, encompassing detection and escape phases. The detection phase involves utilizing an XSS detector, where the input undergoes preprocessing before being fed into the detector. Preprocessing comprises several steps: Positive examples, containing XSS attacks, are de-obfuscated and converted to lowercase. The URL is standardized to `http://', and special characters such as angular brackets in `$<$br$>$' are removed. Tokenization is then applied to the examples using the rules outlined in Table~\ref{tab:tokenization_reference}.

\begin{table}[!h]
    \caption{Tokenization Rules}
    \centering
    \scalebox{1.0}{
        \begin{tabular}{l|l}
        \toprule
        Regular Expression & Object \\
        \midrule
        
        (?x)[\textbackslash w\textbackslash.]+?\textbackslash( &   Javascript function\\
        ''\textbackslash w+?'' &  Content within double quotes\\
        '\textbackslash w+?' &  Content within single quotes\\
        http://\textbackslash w+ &  URLs\\
        $<$\textbackslash w+$>$ &  Opening tags \\
        $<$/\textbackslash w+$>$ &  Termination tags \\
        \textbackslash b\textbackslash w+= & Attributes \\
        (?$<$=\textbackslash ()\textbackslash S+(?=\textbackslash )) & Content within parentheses\\
        
        \bottomrule
        \end{tabular}
    }
    \label{tab:tokenization_reference}
\end{table}

Tokenization converts each input example into a sequence of tokens. The 10\% most frequent tokens are chosen for the vocabulary, while the remaining tokens are replaced with `None'. Subsequently, a Word2Vec model is trained, representing each token as a 32-dimensional vector. The length of each example is standardized to 200 words, discarding excess words and padding shorter examples with 0. The resulting vectors are then fed into the detection models, as shown in Figure~\ref{fig:reference_method}.

The escape phase involves crafting adversarial examples using RL. The idea is to train an agent to generate these examples effectively. The action space is carefully defined as a set of modification operations that can be applied to a malicious example while preserving its inherent characteristics. Table~\ref{tab:actions_reference} provides a comprehensive list of these possible actions.

\begin{table}[!h]
\caption{List of Actions} \label{tab:actions_reference}
\centering
\scalebox {0.7} {
\begin{tabular}{llll}
    \toprule
    
    A1) & Add " \&\#14" before "javascript" & A15) & Replace "(" and ") with "`" \\
    A2) & Mixed case HTML attributes & A16) & Encode data protocol with Base64 \\
    A3) & Replace spaces with ”/”, ”\%0A” or ”\%0D” & A17) & Remove the quotation marks \\
    A4) & Mixed case HTML tags & A18) & Unicode encoding for JS code \\
    A5) & Remove the closing symbol of the single tags & A19) & HTML entity encoding for ”javascript” \\
    A6) & Add ”\&NewLine;” to ”javascript” & A20) & Replace ”$>$” of single label with ”$<$” \\
    A7) & Add ”\&\#x09” to ”javascript” & A21) & Replace ”alert” with ”top[’al’ + ’ert’](1)” \\
    A8) & HTML entity encoding for JS code (hexadecimal) & A22) & Replace ”alert” with ”top[8680439..toString(30)](1)” \\
    A9) & Double write HTML tags & A23) & Add interference string before the example \\
    A10) & Replace ”http://” with ”//” & A24) & Add comment into tags \\
    A11) & HTML entity encoding for JS code (decimal) & A25) & ”vbscript” replaces ”javascript” \\
    A12) & Add ”\&colon;” to ”javascript” & A26) & Inject empty byte ”\%00” into tags \\
    A13) & Add ”\&Tab;” to ”javascript” & A27) & Replace ”alert” with ”top[/al/.source + /ert/. source](1)” \\
    A14) & Add string ”/drfv/” after the script tag & \\
    \bottomrule
\end{tabular}

}

\label{tab:actions}
\end{table}

The state space includes the historical record of actions taken by the agent. During each step, the agent chooses an action, updates its state accordingly, and then submits the transformed examples to the detection model. The agent receives a reward of 10 if the examples successfully evade detection, and a penalty of -1 if they are detected. This process continues iteratively until the agent either successfully bypasses the detection mechanism or reaches the maximum allowed number of steps.

\subsection{Experiments}

The training set for XSS detection models contains around 90,000 examples, collected from XSSed~\cite{kf} and Alexa~\cite{Cooper}.
This dataset is not publicly available and we had no way to craft it. To solve this problem, we used a publicly available dataset, as discussed in Section~\ref{sec:method}.
The authors of the reference work used the same dataset as \cite{wang2022black} to train the adversarial model.
They trained MLP, LSTM, and CNN as detectors, and they considered also two commercial XSS detection systems, named Safedog~\cite{Safedog} and XSSChop ~\cite{Chaitin}.

Several metrics were employed to assess the performance of the detectors: True Positive (TP) represents a correctly identified XSS example, False Positive (FP) indicates a benign example wrongly classified as malicious, True Negative (TN) denotes a correctly identified benign example, and False Negative (FN) represents an XSS example wrongly classified as benign. Then, some derived metrics are defined as follows: \textit{Accuracy} measures the proportion of correctly predicted examples (both malicious and benign) among the total. \textit{Precision} calculates the ratio of correctly predicted malicious examples to all predicted malicious examples. \textit{Recall} determines the ratio of correctly predicted malicious examples to all actual malicious examples. \textit{F1-Score} is the geometric mean of Precision and Recall, aiming for high values of both.

When evaluating a adversarial attack, the authors focused on detection and escape rates. \textit{Detection Rate} (DR) represents the ratio of malicious examples still detected by the XSS detection model, indicating the model's ability to defend against adversarial examples:
\begin{equation}
\label{eqn:6}
 DR = \frac{Number\ of\ malicious\ examples\ detected }{Total\ number\ of\ adversarial\ examples }
\end{equation}

\noindent
\textit{Escape Rate} (ER) refers to the percentage of malicious examples that go undetected and are recognized as benign by the detector:
\begin{equation}
\label{eqn:7}
 ER =\frac{Number\ of\ malicious\ examples\ undetected }{Total\ number\ of\ adversarial\ examples }
\end{equation}

\new{The results presented in the reference work are summarized and analyzed in \ref{appendix_a}}

\subsection{Identified Threats to Validity}

The first threat to validity stems from the lack of validation of adversarial examples and their properties. The authors did not employ any strategy to ensure that the applied transformations preserve the semantic integrity of the examples. This omission raises concerns about the validity of the modified payloads. The second issue is related to the preprocessing and vocabulary construction. The initial dataset lacks tokens produced by the proposed payload transformations, or in some cases such tokens are rare in the dataset, meaning that new tokens resulting from the RL agent's actions are likely to fall outside the top 10\% of considered tokens. Consequently, these tokens will be replaced with `None', causing semantic changes that may invalidate the attack. This issue could lead to an inflated Escape Rate (ER) due to the disruption of the payload's semantics during preprocessing, making it challenging to assess the models' true detection capabilities. The final threat concerns the unavailability of crucial components of the reference work. The code, datasets, and models are not accessible, hindering further analysis and replication of the reported results. This lack of transparency impedes the progress of research in this area, as it becomes difficult to build upon and extend the original findings. We have contacted the original authors asking them for code, datasets and models, but they never replied.

We summarize the three identified threats as follows:
\begin{itemize}
    \item \textbf{TH1. Lack of validation of the actions}: The lack of validation for the applied actions raises questions about the semantic validity of modified payloads, potentially affecting the integrity of the examples.
    
    \item \textbf{TH2. Lack of validation of the preprocessed payload}: The preprocessing pipeline requires validation to ensure that the preprocessed payload maintains its semantic validity, which is crucial for accurate evaluation.
    
    \item \textbf{TH3. Lack of reproducibility}: The unavailability of experimental details, including code, datasets, and models, hinders reproducibility and limits the ability to extend and build upon the research, impeding further advancements in the field.
\end{itemize}

\section{Methodology}
\label{sec:method}

The main idea of this paper is to introduce an XSS Oracle.
As a first step, we demonstrate the Oracle's usefulness in assessing the validity of payloads and their potential impact. Furthermore, the Oracle can aid in developing a robust defense model, which allows an accurate evaluation of the performance of the approach proposed by Chen et al.~\cite{CHEN2022102831}.

\subsection{XSS Oracle}

\begin{figure}[!h]
  \centering
  \includegraphics[width=0.95\columnwidth]{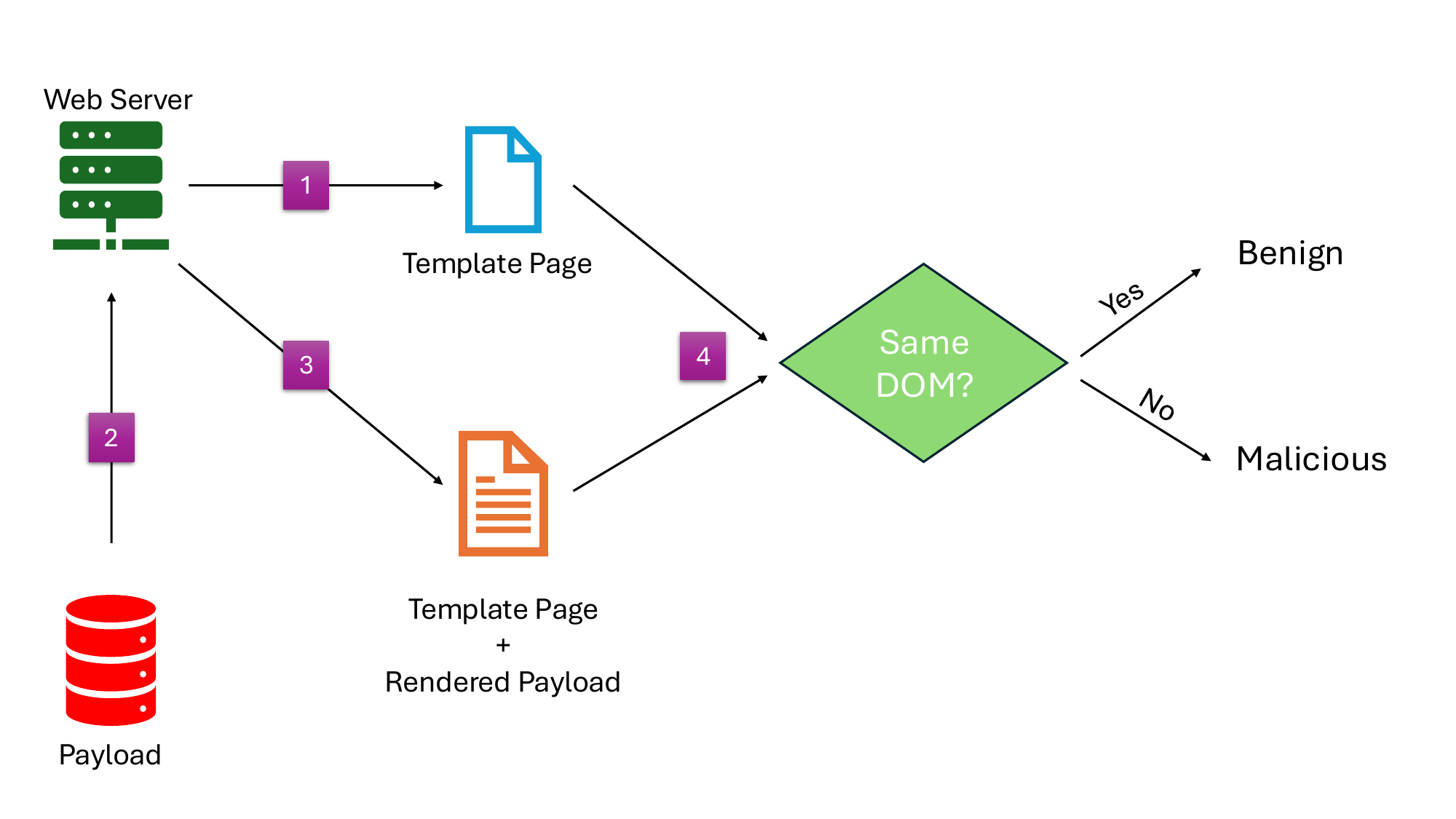}
  
  \caption{Workflow of the XSS Oracle. A Payload is rendered in a known template of a Web Page and the DOMs of the two pages are compared.}\label{fig:Oracle}
\end{figure}

Based on the presence of an XSS attack inside of the payload of an HTTP request, we can consider two types of payload: `Benign' and `Malicious'.
Benign payloads do not alter the DOM structure when executed, while malicious payloads cause changes in the DOM, potentially affecting the browser environment. As outlined in Figure~\ref{fig:Oracle}, we utilize the Oracle to mimic payload execution, observing the DOM of a template page rendered by a web server. The server accepts the payload as a parameter and incorporates its elements into the template. The Oracle then examines the DOM of the new page. If any differences are detected, the payload is labelled as Malicious; otherwise, it is classified as Benign.

\subsection{Metrics}
\label{sec:metrics}
TH1 and TH2 arise from the agent's modifications and preprocessing of the payload, which could alter the characteristics of the XSS attack. To address this, we employ an XSS Oracle that assesses the integrity of the attack properties, introducing the metric Ruin Rate (RR).
Let us consider a set of payloads labeled as Malicious, denoted as $M=\{m_1, m_2, \ldots\}$. This set $M$ is generic and can include malicious samples from the original dataset or those generated by the XSS adversarial method. 

How to structure the different sets for the evaluation of the different threats to validity will be discussed in Section~\ref{sec:empirical_study}.

We define a function $O(p)$ that, for any payload $p$, returns 1 if the Oracle classifies $p$ as Malicious and 0 otherwise. For any set $M$,  RR can be calculated as:

\begin{equation}
    RR(M) = 1 - \frac{\sum_{m \in M}{O(m)}}{|M|}
\end{equation}

If $M$ contains samples from the original dataset, a non-zero $RR(M)$ indicates mislabeled examples. Conversely, if $M$ consists of adversarial examples derived from an original set with $RR=0$, a non-zero $RR(M)$ points to an adversarial process that has compromised the attack's properties.

TH1 and TH2 are also potentially related to an anomalous number of Out-Of-Vocabulary (OOV) tokens in the array fed to the detection model.
We introduce a second metric, called OOV-Rate ($OR$), to evaluate this aspect.
For an array of tokens $V$, $OR(V)$ is the number of the `None' tokens present inside $V$ (these represent OOV tokens) over the length of $V$:
\begin{equation}
    OR(V) = \frac{\sum_{v \in V}{OOV(v)}}{|V|}
\end{equation}

\noindent
where  function $OOV(v)$ is 1 if $v$ = \textit{None}, 0 otherwise.


\section{Empirical Study}
\label{sec:empirical_study}

This section presents the replication of the experiments in the reference work~\cite{CHEN2022102831}, highlighting the deviations from the results reported in the original paper. We introduce specific research questions for the replication and for the extension study, and we describe the experiments conducted to extend the reference work.

\subsection{Replication Study}
\label{sec:replication_study}

In our replication study, we aim to closely follow the methodology of the reference work. However, some differences are worth noting and justifying. The dataset used to train the detectors was not publicly available, so we utilized an alternative dataset.\footnote{\url{https://github.com/fmereani/Cross-Site-Scripting-XSS/blob/master/XSSDataSets/Payloads.csv}}
This employed dataset is a well-known one~\cite{mereani2018howe} containing more than 15,000 Malicious and Benign payloads. \new{This dataset is built by collecting real-life attacks.} The dataset for training the adversarial agent was partially available but had a different structure compared to the one employed in the reference work. The reference work does not adequately describe the correct payload structure, as the examples only consider parameters, while some steps mention filtering applied to the URL, suggesting the payload should be the entire HTTP request. Preliminary experiments revealed that datasets with varying structures encountered out-of-vocabulary issues even before the agent's actions were applied.
In particular, when one dataset is used to create the vocabulary and to train a detector, and the other one is simply tested against it,  RR is very high even before training an adversarial agent ($RR>60\%$).

To isolate the identified threats to validity and mitigate any data structure-related problems, we divided the selected dataset into two parts: one for training the detectors and the other for training the adversarial agents, excluding Benign examples. This setup makes adversarial attacks more challenging, as the examples are closer to those used for detector training \newt{since they belong to the same distribution, thus they are more prone to presenting the same patterns}. Consequently, a high RR in this context would indicate the significance of the threats TH1 and TH2, because of the alignment between detector's and adversarial agent's training sets. 

The dataset was pre-filtered by the Oracle to ensure accurate labelling. After  pre-filtering, the most representative class (Benign) was undersampled to ensure class balance. We computed the Ruin Rate of the original payload before and after preprocessing, resulting in a RR = 0\%, confirming that all samples were initially valid according to our Oracle. Furthermore, the dataset was split into train, validation, and test sets. Table~\ref{tab:dataset} shows such details.

\begin{table}[!h]
\caption{Dataset split between detector and adversarial agent, and then into train, validation and test sets} \label{tab:dataset}
\centering

\begin{tabular}{c|ccc|ccc}
    \toprule
    Label &\multicolumn{3}{c|}{Detectors} & \multicolumn{3}{c}{Adversarial Agent}\\ & Train Set & Val. Set  & Test Set  & Train Set & Val. Set  & Test Set   \\
    \midrule
    Benign & 2,884   & 721 & 902 & 0 & 0 & 0 \\
    Malicious & 2,883   & 721 & 901 & 2,883 & 712 & 901 \\
    \bottomrule
\end{tabular}

\end{table}

For the detection models, we used a different  activation function in the output layer compared to the reference work. The softmax function used in the reference is more appropriate for multiclass classification, whereas our binary classification problem (Benign vs. Malicious) is better suited to the sigmoidal function. Furthermore, we employed only CNN, MLP, and LSTM as detectors, as Safedog and XSSChop are not publicly available. These models were trained for 150 epochs with early stopping (patience of 10 epochs), an embedding dimension of 8, a learning rate of $10^{-3}$, and a stochastic gradient descent optimizer.

In contrast to the reference work, which used the SAC algorithm~\cite{haarnoja2018soft} for agent training, we opted for the PPO algorithm~\cite{schulman2017proximal}. This decision was made because we utilized the Stable Baselines library in Python for the Reinforcement Learning model implementation. The SAC algorithm implementation in this library is designed for a continuous action space, which does not align with our discrete action space, comprising discrete actions for mutating the attack payload. Therefore, we chose the PPO algorithm implementation, which  handles a discrete action space.\footnote{\url{https://stable-baselines3.readthedocs.io/en/master/modules/sac.html}}

The performance of our detection models is similar to the reference work, achieving near-perfect metric scores as presented in Table~\ref{tab:detectors}. 

\begin{table}[!h]
\caption{Performance of the XSS detection models of the replication study.} \label{tab:detectors}
\centering
\scalebox {1.0} {
\begin{tabular}{c|cccc}
    \toprule
    Detector & Precision & Recall  & Accuracy & F1      \\
    \midrule
    MLP        & 99.67\%   & 100.0\% & 99.83\%  & 99.83\% \\
    LSTM       & 99.67\%   & 100.0\% & 99.83\%  & 99.83\% \\
    CNN        & 99.67\%   & 100.0\% & 99.83\%  & 99.83\% \\
    \bottomrule
\end{tabular}
}

\end{table}

\subsection{Research Questions (RQs)}


\noindent The \new{first} two RQs focus on evaluating the significance of TH1 and TH2:
\begin{itemize}
    \item \textbf{\new{RQ1}. Evaluation of TH1}: \textit{Does the lack of validation of the actions pose a threat to validity?}
    \item \textbf{\new{RQ2}. Evaluation of TH2}: \textit{Does the lack of validation of the preprocessed payload pose a threat to validity?}
\end{itemize}

\noindent The final RQ extends this study by re-examining the performance of the method introduced in the reference work after addressing the identified threats to validity.
\begin{itemize}
    \item \textbf{\new{RQ3}. Extension Study}: \textit{How does the reference method perform once the identified threats are mitigated?}
\end{itemize}

\subsection{Implementation}
\begin{figure}[!ht]
  \centering
  \includegraphics[width=0.95\textwidth]{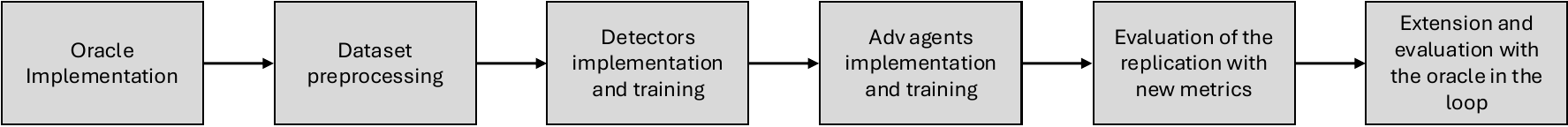}
  \caption{\new{Pipeline of the implementation and experiments, showing the main steps we performed for the replication and extensions study.}} \label{fig:pipeline}
\end{figure}

\new{The details of the implementation were not available (see threat TH3). Hence, we proceeded with the development of a framework that we considered a good fitting of the requirements in the reference work, at the same time compatible with our own extension.}
Our experimental framework was implemented using Python 3.11. The DL library used to implement the models is PyTorch 2.2.1. \new{The detectors are trained using a pre-processing pipeline that closely follows the instructions of the reference work, while the training procedure has been performed after a hyper-parameter tuning procedure.
Figure~\ref{fig:pipeline} shows the main steps performed for the implementation and the execution of the experiments of this study.
}
The RL agent used for generating adversarial attacks is implemented in StableBaselines3 2.3.0.
\new{The agent's actions are implemented following the instructions of the reference work (see Table~\ref{tab:actions_reference} for the comprehensive list of actions that we have implemented), and have been manually evaluated and tested using the Oracle. The reward function $f(x)$ used by the RL agent was also taken from the original work (see  Sec.~\ref{sec:proposed_method}), and is defined as follows for a mutated payload $x$:}

\begin{equation}
\new{
f(x) = 
\begin{cases}
10  ~~~\text{if attack $x$ goes undetected} \\
-1  ~~~\text{if attack $x$ is detected}
\end{cases}
}
\end{equation}

The Oracle is implemented with a Web Server using FastAPI 0.104.0 and Jinja2 3.1.2 to render the template. The DOM is analyzed using BeautifulSoup 0.0.2 and zss 1.2.0.

\subsection{Oracle Integration and Analysis}

\begin{figure}[!ht]
  \centering
  \includegraphics[width=0.9\textwidth]{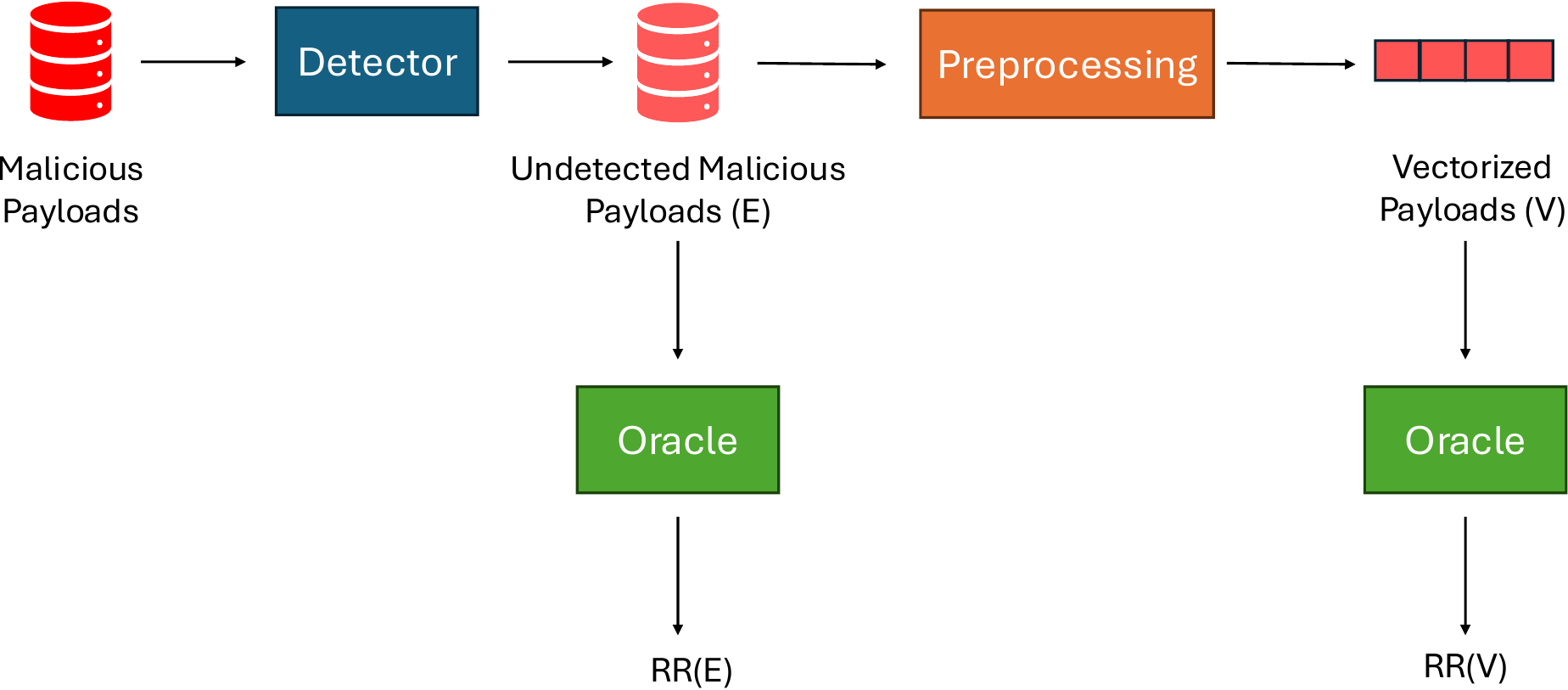}
  \caption{Experimental setup: malicious undetected payloads are fed into the Oracle before and after  preprocessing\label{fig:experiments}. \newt{The Oracle is used to assess whether Undetected Malicious Payloads (E) and their Vectorized representations (V) preserve the properties of an XSS attack, and to compute the corresponding Ruin Rate (RR).}}
\end{figure}

\begin{figure}[!ht]
  \centering
  \includegraphics[width=0.9\textwidth]{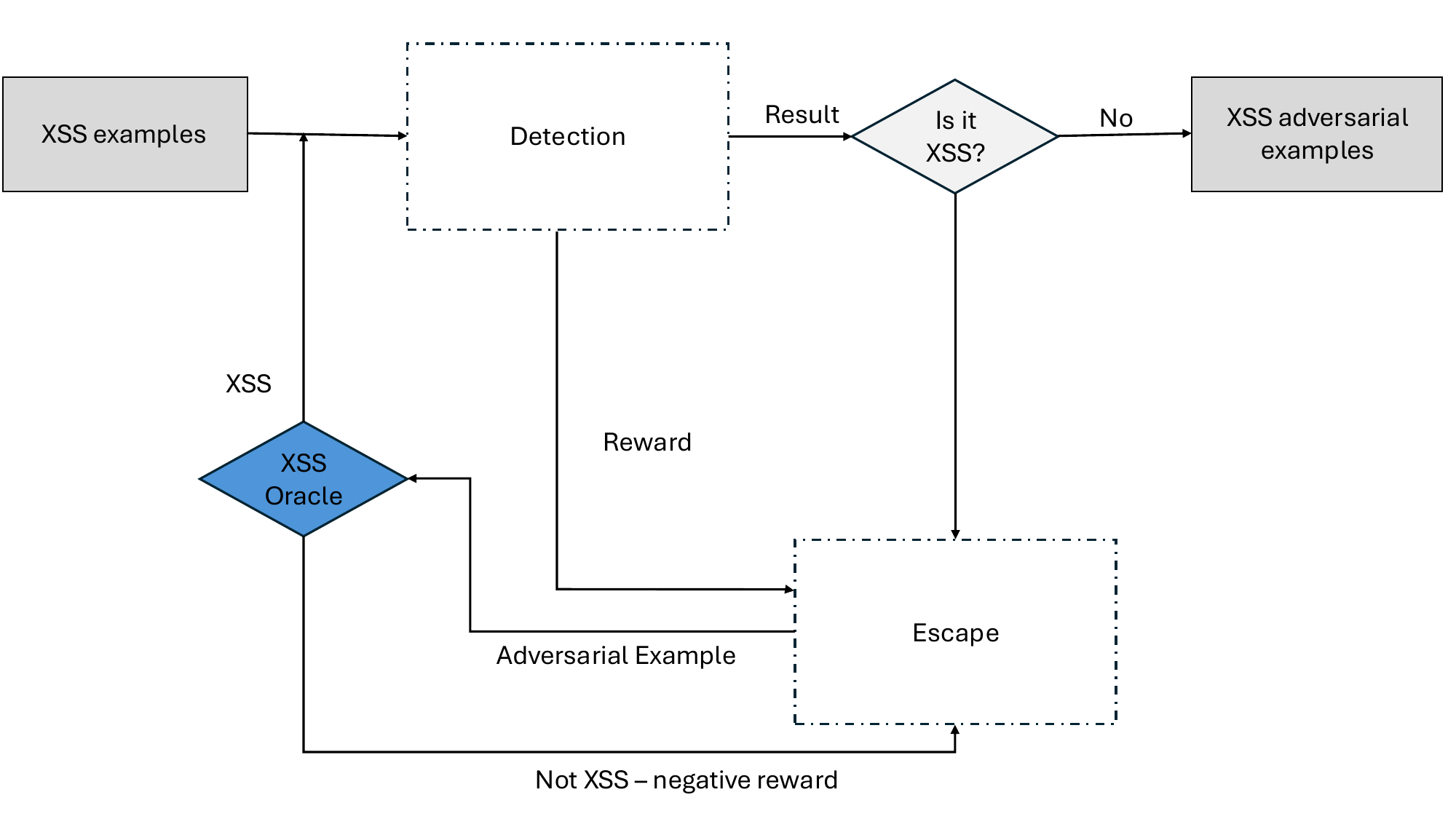}
  \caption{\new{Adversarial approach pipeline with the integration of the XSS Oracle. This is similar to the original one presented in Figure~\ref{fig:reference_method}, with the addition of the Oracle that returns a negative reward if the mutated attack is no longer a valid XSS.}}\label{fig:replication_study_approach}
\end{figure}

As shown in Figure~\ref{fig:experiments}, the Oracle is used in two different stages.
The set of undetected malicious payloads generated by the adversarial model, which represent the examples that contribute to the escape rate of the replication study, named $E$, is directly fed into the Oracle.
The set $E$ is then preprocessed as described in the previous sections, obtaining the set of arrays $V$.
Also, $V$ is fed into the Oracle. 
Thanks to the Oracle, it is possible to evaluate $RR(E)$ and $RR(V)$, which, respectively, represent the answers to \new{RQ1} and \new{RQ2}.
We do not rely only on the Oracle: the analysis of the Ruin Rate for \new{RQ2} is complemented by the analysis of the Out-Of-Vocabulary Rate, that it is not reported in the Figure~\ref{fig:experiments} for simplicity. 
Regarding \new{RQ3}, there is no guarantee that increasing the vocabulary would solve TH2, since the adversarial agent is potentially able to generate new tokens that are out-of-vocabulary regardless of the vocabulary size.
To mitigate this threat-to-validity and to evaluate the real performance of the method, we integrated the Oracle into the training process of the adversarial agent.
The agent's reward is set to -2 if the mutated payload, after  preprocessing, is no longer recognized by the Oracle as an XSS attack.
Otherwise, the reward is the same proposed in the reference work.
\new{The entire adversarial approach pipeline after the introduction of the Oracle in the loop is reported in Figure~\ref{fig:replication_study_approach}.}

\new{
\section{Results of the Replication Study}
\label{sec:rq1}

\begin{table}[!h]
\caption{Escape rate of the original adversarial agent, averaged across 10 training repetitions} \label{tab:rq1}
\centering

\begin{tabular}{ll}
    \toprule
    Detector & Escape Rate \\
    \midrule
    LSTM            & 98.62\%     \\
    MLP             & 99.73\%     \\
    CNN             & 98.25\%     \\
    \bottomrule
\end{tabular}

\end{table}

Before analyzing the results, it is relevant to describe how we mitigate TH3 by replicating the results of Chen et al.~\cite{CHEN2022102831}. We trained ten adversarial agents attacking each considered detection model, to deal with the non-determinism of the training process. Table~\ref{tab:rq1} reports the average of the escape rates obtained by the adversarial agents. These ERs are almost perfect, demonstrating a consistency with the reference work, despite variations in the dataset and training algorithm (see Section~\ref{sec:replication_study}). LSTM's ER is 6.58\%pt\footnote{Percentage points \%pt is the standard unit of measure for differences between percentages (e.g., 80\% is 100\%, or 40\%pt, higher than 40\%).} higher than the reference work, while the MLP and CNN results are identical and slightly lower (0.99\%pt), respectively. This consistency strongly suggests that the differences in the dataset and training algorithm did not significantly impact the overall outcome.
Our replication study successfully reproduced the reference work's results, confirming the effectiveness of the proposed adversarial agents.

}
\section{Results}
\label{sec:results}

Since we successfully replicated the original study, with negligible differences in the results despite the changed dataset and training algorithm, we proceeded to test our hypotheses, trying to explain the reasons for such an amazing performance. We conjectured that the adversarial agents are exploiting vulnerabilities arising from the lack of validation of the actions and due to out of vocabulary tokens produced by preprocessing, which would rendering the other aspects of the algorithm less significant. In fact, any algorithm whose actions produce ineffective or out of vocabulary payloads would achieve a high escape rate, without generating any meaningful attack.


\subsection{\new{RQ1} (Evaluation of TH1)} \label{sec:rq2}

\begin{table}[!h]
\caption{Ruin rates of  set E, averaged across 10 repetitions} \label{tab:rq2}
\centering
\scalebox {1.0} {

\begin{tabular}{ll}
    \toprule
    Detector & $RR(E)$ \\
    \midrule
    LSTM            & 6.34\%     \\
    MLP             & 7.07\%     \\
    CNN             & 6.36\%     \\
    \bottomrule
\end{tabular}

}
\end{table}

We investigate the impact of the lack of action validation by analyzing the ruin rates of the set $E$, which contains all the generated payloads that successfully escaped detection. Table~\ref{tab:rq2} presents the average ruin rates ($RRs$) for each detection model.

Ruin rates are relatively low, ranging from 6.34\% to 7.07\%, indicating that the sequence of actions occasionally disrupts the semantics of the attack. However, this frequency is not high enough to be considered a significant threat to validity.

\begin{tcolorbox}[boxrule=0pt,frame hidden,sharp corners,enhanced,borderline north={1pt}{0pt}{black},borderline south={1pt}{0pt}{black},boxsep=2pt,left=2pt,right=2pt,top=2.5pt,bottom=2pt]
\textbf{Answer to \new{RQ1}}: The lack of validation of the actions is not a severe threat to validity, but it warrants further investigation to improve the detection model's robustness.
\end{tcolorbox}

\subsection{\new{RQ2}  (Evaluation of TH2)}

\label{sec:rq3}
\begin{table}[!h]
\caption{Ruin and OOV rates of the array V, averaged across 10 repetitions}
\label{tab:rq3}
\centering
\scalebox {1.0} {
\begin{tabular}{lll}
    \toprule
    Detector & $RR(V)$ & $OR(V)$ \\
    \midrule
    LSTM            & 97.31\% & 47.49\%    \\
    MLP             & 97.84\% & 44.76\%  \\
    CNN             & 92.57\% & 43.85\%   \\
    \bottomrule
\end{tabular}
}
\end{table}

For each adversarial agent, we collected all generated payloads that bypassed the detector, forming the set $E$. We then preprocessed this set to create the array $V$, and analyzed its ruin rate ($RR(V)$). The second column of Table~\ref{tab:rq3} presents the average ruin rates across ten adversarial agents for each detection model. Ruin rates after preprocessing are notably high (exceeding 90\% in all cases), indicating that preprocessing significantly disrupts the semantics of the XSS attack, posing a concrete threat to the validity of the original empirical study.

To find further explanations, we considered the content of $V$ to assess whether our hypothesis about the introduction of `None' (causing OOV tokens) could be a contributing factor to the performance of the original adversarial agent. 
The third column of Table~\ref{tab:rq3} displays the average OOV rates across the ten adversarial agents for each detection model.  OOV rates are significantly high, exceeding 40\% in all cases, which is notably higher than the OOV rates of non-mutated payloads (around 6\%). This confirms the threat to validity TH3 and suggests that preprocessing is one of the primary reasons for the agent's high escape rate, as adversarial agents learn to exploit OOV tokens as a \textit{shortcut} to escape detection, rather than generating valid XSS payloads that can bypass detection while retaining their semantic integrity after preprocessing.

It is important to notice that the attack described in the reference work remains effective and poses a risk. 
However, in the original setup the preprocessing step makes the attack successful for any detection model. Any attack that trivially introduces OOV tokens is effective by construction in such setup. However, a defender aware of the OOV token issue would implement an additional layer of protection to discard payloads with an unusually high number of OOV tokens, rendering the attack ineffective, which complicates the assessment of the attack's true effectiveness against a wide range of detectors and raise questions about the vulnerabilities of commercial systems like XSSChop or SafeDog to such attacks, because all these detection systems might be in principle unaware of the OOV token problem.

\begin{tcolorbox}[boxrule=0pt,frame hidden,sharp corners,enhanced,borderline north={1pt}{0pt}{black},borderline south={1pt}{0pt}{black},boxsep=2pt,left=2pt,right=2pt,top=2.5pt,bottom=2pt]

\textbf{Answer to \new{RQ2}}: The lack of validation of the preprocessed payload poses a concrete threat to validity. This threat arises from the high ruin and OOV rates observed, indicating that preprocessing disrupts the XSS payload semantics. Adversarial agents exploit this, learning to bypass detection without preserving the payload meaning. This highlights the need for improved payload validation techniques to mitigate the possibility of attack shortcuts due to preprocessing.
\end{tcolorbox}

\subsection{\new{RQ3} (Extension Study)}
\label{sec:rq4}

\begin{table}[!h]
\caption{Escape rates of the adversarial agent with the Oracle included in the training process, averaged across 10 repetitions} \label{tab:rq4}
\centering
\scalebox {1.0} {
\begin{tabular}{ll}
    \toprule
    Detector & Escape Rate \\
    \midrule
    LSTM            & 98.13\%     \\
    MLP             & 97.37\%     \\
    CNN             & 96.89\%     \\
    \bottomrule
\end{tabular}
}
\end{table}

In this RQ, the training process for adversarial agents closely mirrors that used in \new{the replication study} (Section~\ref{sec:rq1}), with a key difference: the Oracle is integrated to calculate a new reward function. The new reward is set to $-2$ if the mutated payload, after  preprocessing, is no longer recognized by the Oracle as an XSS attack. Otherwise, the reward is the same proposed in the reference work. Table~\ref{tab:rq4} reports the average escape rates achieved by the adversarial agent. Despite being very high, these rates are slightly lower than those obtained in \new{the replication study} (see Table~\ref{tab:rq1}).

\begin{table}[!h]
\caption{Ruin and OOV rates of the array V, evaluated for \new{RQ3} and averaged across 10 repetitions}
\label{tab:rq4-rr}
\centering
\scalebox {1.0} {
\begin{tabular}{lll}
    \toprule
    Detector & $RR(V)$ & $OR(V)$ \\
    \midrule
    LSTM            & 0.01\% & 1.73\%    \\
    MLP             & 0.11\% & 2.30\%  \\
    CNN             & 0.08\% & 1.90\%   \\
    \bottomrule
\end{tabular}
}
\end{table}

The second and third columns of Table~\ref{tab:rq4-rr} report the average ruin rates ($RR$) and out-of-vocabulary (OOV) rates ($OR$) across the ten adversarial agents for each detection model. The low values of $RR$ and $OR$ indicate that the integration of the Oracle in the training process effectively mitigates TH3. These results demonstrate that it is feasible to train adversarial agents capable of attacking XSS detection models as proposed in the reference work, without introducing any threats to validity related to preprocessing. In the new setup, the adversarial agent learns to produce payloads that include mostly valid tokens, while being still able to circumvent the detection capabilities of the considered detectors.

\begin{tcolorbox}[boxrule=0pt,frame hidden,sharp corners,enhanced,borderline north={1pt}{0pt}{black},borderline south={1pt}{0pt}{black},boxsep=2pt,left=2pt,right=2pt,top=2.5pt,bottom=2pt]

\textbf{Answer to \new{RQ3}:} Our Oracle-enhanced training method demonstrates the concrete threat of adversarial attacks on XSS detectors. Despite slight performance degradation, it confirms the ability of these attacks to bypass detection without exploiting threats related to the preprocessing.

\end{tcolorbox}

\section{Threats to Validity}
\label{sec:threats}

\noindent \textbf{Internal validity}. 
The training process of the adversarial agents is inherently non-deterministic. To ensure reliability of our findings, we repeated the training process for each agent ten times. For transparency and correctness of implementation, we have made our code publicly available, and we utilized well-known open-source frameworks for our implementation. \newt{The introduction of a customized XSS Oracle also poses a threat to internal validity, as it may be subject to misclassification and depend on the specific templates of the dataset used to construct and evaluate it.}

\noindent \textbf{External validity}.
While the employed dataset may not be exhaustive in representing every type of XSS attack, it is substantial and publicly accessible. Moreover, it has been widely used in previous research, establishing it as a suitable benchmark for evaluating XSS detection methods. \newt{Nevertheless, our results are inherently tied to the specific datasets and detector architectures considered, which may limit their generalizability.}

\noindent \textbf{Construct validity}. We employed standard evaluation metrics in the security domain, including Precision, Recall, Accuracy, and F1-Score, to assess the detectors. For the adversarial agents, we used the escape rate as an evaluation metric, which aligns with the reference work. \new{The Ruin Rate is defined with respect to a single, explicitly implemented XSS Oracle; different oracle designs or interpretations of attack preservation could therefore lead to different measurements.}

\newt{\noindent \textbf{Replication fidelity}. Because the original implementation is not publicly available, our replication necessarily relied on a set of assumptions and exhibits differences from the reference work. In addition, the commercial XSS detectors used in the original study are not accessible. These factors introduce potential threats to replication fidelity.}
\new{
\section{Discussion}
\label{sec:discussion}
This work primarily focuses on identifying and mitigating the threats to validity present in prior research~\cite{CHEN2022102831}. Our extension further leads to the design and adoption of a fairer evaluation framework — one that more accurately reflects the true capability of an attack algorithm to evade detection by a DNN-based detector.
It is important to note, however, that an attack that remains undetected due to disruptive preprocessing still constitutes a successful attack, since it ultimately achieves the attacker’s objective. Therefore, in this section, we discuss how our findings can inform practical defensive strategies.

A key takeaway from our study is the critical role of preprocessing and vocabulary construction. These components deserve as much attention and refinement as the DNN architecture itself, given their substantial impact on the detector’s robustness. We recommend that detectors be systematically evaluated in adversarial setups, where metrics such as the OOV rate can guide defenders in identifying and prioritizing the weakest parts of their detection pipeline for improvement.
In particular, the RR metric primarily exposes weaknesses in the design of the adversarial strategy, offering limited actionable insights for defenders. In contrast, the OOV metric is strongly tied to vulnerabilities in the preprocessing pipeline. Consequently, in adversarial testing scenarios, a low RR coupled with a high OOV signals an urgent need to revisit and possibly redesign the preprocessing stage.

Finally, while our case study targeted a specific work~\cite{CHEN2022102831}, the same threats to validity are likely to affect other state-of-the-art approaches as well. Replicating additional SOTA studies using our proposed metrics would therefore provide a more comprehensive assessment of the robustness and generalizability of such detection systems.
}
\section{Conclusion}
\label{sec:conclusion}

In this paper, we replicated the study proposed by Chen et al.~\cite{CHEN2022102831} and conducted a thorough analysis of potential threats to its validity. After checking whether such potential threats actually affected the results reported in the original study, we presented an extended approach and introduced an extension study to mitigate them. Our findings are similar to those presented in Chen et al.~\cite{CHEN2022102831}, but with a crucial difference: we eliminated the threats to their validity. This achievement allows us to propose a more effective method that directly attacks the detectors themselves, rather than relying on potential vulnerabilities in the preprocessing pipeline, associated with the generation of out of vocabulary tokens. Furthermore, our approach enhances transparency in the evaluation process, as we make code, datasets and results publicly available to all researchers in the field.

\section*{Data Availability}
\label{sec:avail}
The implementations, source code, data, and experimental results are publicly available in a GitHub repository\footnote{\url{https://github.com/GianlucaMaragliano/Adversarial_RL_XSS}}.

\section*{Acknowledgement}
This work is funded by the European Union's Horizon Europe research and innovation programme under the project Sec4AI4Sec, grant agreement No 101120393.

\bibliographystyle{elsarticle-harv}
\bibliography{bibliography.bib}

\appendix
\section{Results Reported in the Reference Work}
\label{appendix_a}
The performance of the  XSS detectors is reported in Table~\ref{tab:detectors_reference}, showing the usefulness of the considered models in detecting XSS attacks with almost perfect results.

\begin{table}[!h]
\caption{Performance of the XSS detection models considered in the reference work}
\label{tab:detectors_reference}
\centering
\scalebox {1.0} {
\begin{tabular}{c|cccc}
    \toprule
    Detector & Precision & Recall  & Accuracy & F1      \\
    \midrule
    MLP        & 99.92\%   & 98.00\% & 99.61\%  & 98.96\% \\
    LSTM       & 99.97\%   & 98.35\% & 99.65\%  & 99.06\% \\
    CNN        & 99.85\%   & 98.90\% & 99.76\%  & 99.38\% \\
    XSSChop    & 99.61\%   & 98.25\% & 99.14\%  & 98.93\% \\
    SafeDog    & 100.00\%  & 96.16\% & 98.47\%  & 98.05\% \\ 
    
    \bottomrule
\end{tabular}
}

\end{table}

Regarding the adversarial attacks, Chen et al.~\cite{CHEN2022102831} computed the ER against all the detectors, showing almost perfect ERs, as reported in Table~\ref{tab:adversarial_reference}. These results were the starting point for our replication study, which was initially triggered by the astonishingly high performance exhibited by the proposed RL-based attack generator. We wanted to understand in depth the reasons for such amazing success.

\begin{table}[!h]
\caption{Results of adversarial attacks in the reference work}
\label{tab:adversarial_reference}
\centering
\scalebox {1.0} {
\begin{tabular}{c|c}
    \toprule
    Detector & Escape Rate (ER) \\
    \midrule
    MLP             & 99.73\%     \\
    LSTM            & 92.04\%     \\
    CNN             & 99.24\%     \\
    XSSChop         & 98.46\%     \\
    SafeDog         & 99.95\%     \\ 
    \bottomrule
\end{tabular}
}
\end{table}

\end{document}